\documentclass[aps,twocolumn,prd,superscriptaddress,preprintnumbers,showpacs]{revtex4}
\usepackage{graphicx}
\usepackage{bm}
\usepackage{amsmath}
\usepackage{amssymb}
\usepackage{amsthm}

\newcommand{\be}{\begin{equation}}
\newcommand{\ee}{\end{equation}}
\newcommand{\bea}{\begin{eqnarray}}
\newcommand{\eea}{\end{eqnarray}}
\newcommand{\bdm}{\begin{displaymath}}
\newcommand{\edm}{\end{displaymath}}
\newcommand{\beas}{\begin{eqnarray*}}
\newcommand{\eeas}{\end{eqnarray*}}

\begin{document}
\title{Variation of fundamental parameters and dark energy. A principal component approach}

\author{L. Amendola}
\email[]{l.amendola@thphys.uni-heidelberg.de}
\affiliation{Institut f\"ur Theoretische Physik, Universit\"at Heidelberg, Philosophenweg 16, 69120 Heidelberg, Germany}
\author{A.C.O. Leite}  
\email[]{up090308020@alunos.fc.up.pt}
\affiliation{Faculdade de Ci\^encias, Universidade do Porto, Rua do Campo Alegre, 4150-007 Porto, Portugal}
\affiliation{Centro de Astrof\'{\i}sica, Universidade do Porto, Rua das Estrelas, 4150-762 Porto, Portugal}
\author{C.J.A.P. Martins}
\email[]{Carlos.Martins@astro.up.pt}
\affiliation{Centro de Astrof\'{\i}sica, Universidade do Porto, Rua das Estrelas, 4150-762 Porto, Portugal}
\author{N.J. Nunes}
\email[]{n.nunes@thphys.uni-heidelberg.de}
\affiliation{Institut f\"ur Theoretische Physik, Universit\"at Heidelberg, Philosophenweg 16, 69120 Heidelberg, Germany}
\author{P.O.J. Pedrosa}
\email[]{ppedrosa@alunos.fc.up.pt}
\affiliation{Faculdade de Ci\^encias, Universidade do Porto, Rua do Campo Alegre, 4150-007 Porto, Portugal}
\affiliation{Centro de Astrof\'{\i}sica, Universidade do Porto, Rua das Estrelas, 4150-762 Porto, Portugal}
\author{A. Seganti}
\affiliation{Dipartimento di Fisica,  Universit\`a di Roma "La Sapienza", P.le Aldo
Moro 2, 00185 Roma, Italy}

\date{1 October 2011}

\begin{abstract}
We discuss methods based on Principal Component Analysis to constrain the dark energy equation of state using a combination of Type Ia supernovae at low redshift and spectroscopic measurements of varying fundamental couplings at higher redshifts. 
We discuss the performance of this method when future better-quality datasets are available, focusing on two forthcoming ESO spectrographs -- ESPRESSO for the VLT and CODEX for the E-ELT -- 
which include these measurements as a key part of their science cases. These can realize the prospect of a detailed characterization of dark energy properties almost all the way up to redshift 4. 
\end{abstract}

\keywords{Cosmology: Theory, Dark Energy, Variation of alpha}
\pacs{98.80.-k,98.80.Jk}
\maketitle

\section{Introduction}
\label{introduction}

Cosmology has recently entered a precision, data-driven era. The availability of 
ever larger, higher-quality datasets has led to the so-called concordance model. 
This is a remarkably simple model (with a small number of free parameters) which 
provides a very good fit to the existing data. However, there is a price to pay for 
this success: the data suggests that $96\%$ of the contents of the universe is in a 
still unknown form. This is often called the dark component of the universe. 
Whatever this may be, all the evidence suggests that it is not composed by the protons, neutrons 
and electrons that we are familiar with, but it must be in some form never seen in the 
laboratory.

Current best estimates suggest that this dark component is in fact a combination of 
two distinct components. The first is called dark matter (making about $23\%$ of the 
universe) and it is clustered in large-scale structures like galaxies. The second, 
which has gravitational properties very similar to those of the cosmological 
constant first proposed by Einstein, is called dark energy and currently dominating 
the universe, with about $73\%$ of the density of the universe

Understanding what constitutes this dark energy is one of the most important 
problems of modern cosmology. In particular, we would like to find out if it is 
indeed a cosmological constant \cite{Carroll:2000fy}, since there are many possible 
alternatives \cite{Copeland:2006wr}. These alternative models often involve scalar 
fields, an example of which is the Higgs field which the LHC is searching for. A 
further alternative are the so-called modified gravity models (for a review see e.g. \cite{2010deto.book.....A,Clifton:2011jh}), in which the 
large-scale behavior of the gravitational interaction is different from that 
predicted by Einstein's gravity.

The main difference between the cosmological constant and the models involving 
scalar fields (which are often collectively called dynamical dark energy models) is 
that in the first case the density of dark energy is always constant (it does not 
get diluted by the expansion of the universe) while in the second one the dark 
energy density does change. One way to distinguish the two possibilities is to find 
ways to measure the dark energy density at several epochs in the universe.

Astrophysical measurements of nature's dimensionless fundamental coupling constants 
\cite{Martins:2002fm,GarciaBerro:2007ir,Uzan:2010pm,martins} can be used to study the properties of dark 
energy, either by themselves or in combination with other cosmological datasets 
(such as Type Ia supernovas and the cosmic microwave background). The concept behind 
this method is described in \cite{Nunes:2003ff,Avelino:2006gc,Avelino:2008dc,2009MmSAI..80..785N} (see also \cite{Parkinson:2003kf}). 
It complements other methods due to its large redshift lever arm and the fact that 
these measurements can be done from ground-based telescopes.

Here we revisit this issue and forecast the number of well constrained modes of the dark energy equation of state parameter using a combination of supernovae data and measurements of varying fundamental couplings at high 
redshift.
We will test and validate the reconstruction pipelines by applying them to simulated datasets representative of forthcoming high-quality measurements.
This is particularly relevant for the measurements of varying couplings: 
the existing spectroscopic measurements of the fine-structure constant $\alpha$ 
\cite{Murphy:2003hw,Murphy:2003mi,Srianand:2007qk,Webb:2010hc} and the proton to 
electron mass ratio $\mu$ \cite{Reinhold:2006zn,King:2008ud,Thompson} typically come 
from observations that were not gathered with this purpose in mind, and therefore 
may be vulnerable to considerable uncertainties that are not always easy to quantify.

We will quantify by employing Principal Component Analysis (PCA) (see e.g. \cite{Huterer:2002hy,2007PhRvD..75j3003A} ) 
what improvements will result from the availability of spectrographs
like ESPRESSO (for the VLT) and CODEX (for the E-ELT), which include  measurements
of $\alpha$ and $\mu$                             
as a key part of their science cases. For this purpose we will assume, in either case,
several scenarios for the datasets of fine-structure constant measurements, which will
differ in the number and precision of the measurements.

We should stress that we are not proposing new data analysis techniques. We are extending PCA methods available in the published literature (for type Ia supernovae, lensing and several other contexts in cosmology) and studying the feasibility of applying them to a new type of datasets (astrophysical measurements of varying couplings).

Briefly, the plan of the paper is as follows. In Sec. \ref{methods} we briefly review
the PCA technique, as it applies to our present purposes.
In particular we discuss possible strategies for choosing the number of components and
describe how to build the relevant Fisher matrix for the case of varying couplings. In
Sec. \ref{results} we apply our methods to several scenarios relevant for ESPRESSO and
CODEX. 
In Sec.~\ref{reconstruction} we attempt a reconstruction of the equation of state parameter $w(z)$ using a truncation of the high frequency modes.
Finally in Sec. \ref{concls} we present some conclusions and highlight future
work.

\section{\label{methods}Principal component analysis}

PCA is a non-parametric method for constraining the dark energy equation of state. In assessing its performance, one should not compare it to parametric methods. Indeed, no such comparison is possible (even in principle), since the two methods are addressing different questions. Instead one should compare with another non-parametric reconstruction, and for our purposes with varying couplings the type Ia supernovae provide a relevant comparison.

A key advantage of PCA techniques is that they allow one to infer which and how
many parameters can 
be most accurately determined with a given experiment. Instead of assuming a
parametrization for the relevant observable with a set of parameters born of our
theoretical prejudices, the PCA method leaves the issue of finding the best
parametrization to be decided by the data itself. This is particular useful in
the case of dark energy where, apart from the case of a cosmological constant,
one would be hard pressed to find solid motivations for particular
parametrizations.

In Refs.~\cite{Huterer:2002hy,2007PhRvD..75j3003A} the PCA approach was applied to the use of
supernova data to constrain the dark energy equation of state, $w(z)$. We start
by recalling some of their formalism, which we will then generalize for
measurements of the fine-structure constant $\alpha$. In general one can divide
the redshift range of the survey into $N$ bins such that in 
bin $i$ the equation of state parameter takes the value $w_i$, 
\begin{equation} 
w(z) = \sum_{i = 1}^N w_i \theta_i(z) \,. 
\end{equation} 
Another way of saying this 
is that $w(z)$ is expanded in the basis $\theta_i$, with $\theta_1 =
(1,0,0,...)$, 
$\theta_2 = (0,1,0,...)$, etc.

The precision on the measurement of $w_i$ can be inferred from the Fisher matrix
of the parameters $w_i$, specifically from $\sqrt{(F^{-1})_{ii}}$, and increases
for larger redshift. One can however find a basis in which all the parameters
are 
uncorrelated. This can be done by simply diagonalizing the Fisher matrix such
that 
$F = W^T \Lambda W$ where $\Lambda$ is diagonal and the rows of $W$ are the 
eigenvectors $e_i(z)$ or the principal components. These define the new basis in 
which the new coefficients $\alpha_i$ are uncorrelated and now we can write 
\begin{equation} 
\label{recw} w(z) = \sum_{i = 1}^N \alpha_i e_i(z) \,. 
\end{equation} 
The diagonal elements of $\Lambda$ are the eigenvalues $\lambda_i$ (ordered 
from largest to smallest)
and define the variance of the new parameters, $\sigma^2(\alpha_i) = 1/\lambda_i$.

\subsection{Building the Fisher matrix}

We will consider the standard class of models for which the variation of the
fine-structure constant $\alpha$ is linearly proportional to the displacement of
a scalar field, and further assume that this field is a quintessence type field,
i.e.  responsible for the current acceleration of the Universe \cite{Dvali:2001dd,Chiba:2001er,Anchordoqui:2003ij,Copeland:2003cv,Marra:2005yt,Dent:2008vd,Bento:2008cn}. We take the
coupling between the scalar field and electromagnetism to be 
\be 
{\cal L}_{\phi F} = - \frac{1}{4} B_F(\phi) F_{\mu\nu}F^{\mu\nu} ,
\ee 
where the gauge kinetic function $B_F(\phi)$ is linear, 
\be 
B_F(\phi) = 1- \zeta \kappa (\phi-\phi_0) ,\label{coupling}
\ee 
$\kappa^2=8\pi G$ and $\zeta$ is a constant to be marginalized 
over. This can be seen as the first term of a Taylor expansion, and should be a good 
approximation if the field is slowly varying at low redshift. Then, the evolution of $\alpha$ is given by 
\begin{equation} 
\frac{\Delta \alpha}{\alpha} \equiv 
\frac{\alpha-\alpha_0}{\alpha_0} = \zeta \kappa (\phi-\phi_0) \,. 
\end{equation} 
For a flat Friedmann-Robertson-Walker Universe with a canonical scalar field,
$\dot{\phi}^2 = (1+w(z))\rho_\phi$, hence, for a given dependence of the
equation of state parameter $w(z)$ with redshift, the scalar field evolves as 
\begin{equation} 
\phi(z)-\phi_0 = \frac{\sqrt{3}}{\kappa} 
\int_0^z \sqrt{1+w(z)} \left(1+ \frac{\rho_m}{\rho_\phi}\right)^{-1/2} 
\frac{dz}{1+z} . 
\end{equation} 
where we have chosen the positive root of the solution.

Let us construct the Likelihood function for a generic observable $m(z_i,w_i,c)= 
\mu(z_i,w_i) + c$. For the present purposes this can be the apparent magnitude of a supernova, in which case 
\begin{equation}
\mu = 5 \log (H_0 d_L)\,,\qquad c = M +25 - 5 \log H_0
\end{equation}
or it can be connected to the relative variation of $\alpha$ obtained with quasar absorption spectra, for which
\begin{equation}
\mu = \ln[\kappa(\phi-\phi_0)]\,,\qquad c = \ln \zeta\,.
\end{equation}
Then we find
\begin{equation} L(w^i,M) 
\propto \exp \left[ -\frac{1}{2} \sum_{i,j = 1}^N (m-m_F)_i C_{ij}^{-1} (m-m_F)_j 
\right] . 
\end{equation} 
where $m_F$ means $m$ evaluated at the fiducial values of 
the parameters, $m_F = m_F(z_i,w_i^F,c^F)$ and $C^{-1}$ is the inverse of the 
correlation matrix of the data.

Defining $\beta = c-c^F$, and integrating the likelihood in $\beta$, we obtain the 
marginalized likelihood 
\begin{widetext}
\begin{eqnarray} L(w_i) \equiv \int_{-\infty}^\infty 
L(w_i,\beta) d \beta \nonumber = \sqrt{\frac{2\pi}{A}} \exp\left[ -\frac{1}{2} 
\sum_{i,j = 1}^N (\mu-\mu_F)_i D_{ij}^{-1} (\mu-\mu_F)_j \right] 
\end{eqnarray} %
where $A = \sum_{i,j} C_{i,j}^{-1}$ and 
\begin{equation} D_{ij}^{-1} = C_{ij}^{-1} 
- \frac{1}{A} \sum_{k,l=1}^N C_{kj}^{-1} C_{li}^{-1} . 
\end{equation}
The Fisher matrix can be obtained by approximating  $ L(w_i)$ as a Gaussian in the 
theoretical parameters $w_i$ (the equation of state in each bin) centered around the fiducial model,
and taking the inverse of the resulting correlation function.
The Fisher matrix turns  out to be
\begin{eqnarray} F_{kl} \equiv \left. - 
\frac{\partial^2 \ln L}{\partial w_k \partial w_l}\right|_{w^F} \nonumber = 
\sum_{i,j = 1}^N \left. \frac{\partial \mu(z_i)}{\partial w_k}\right|_{w^F} 
D_{ij}^{-1} \left. \frac{\partial \mu(z_j)}{\partial w_l}\right|_{w^F} , 
\end{eqnarray} 
\end{widetext}
where the derivatives are evaluated at the fiducial values of the 
parameters.

We will consider three fiducial forms of the equation of state parameter:
\begin{eqnarray} 
\label{fid1}
w^F(z) &=& -0.9 ,\\
\label{fid2}
w^F(z) &=& -0.5 + 0.5 \tanh(z-1.5) ,\\ 
\label{fid3}
w^F(z) &=&  -0.9 + 1.3 \exp\left(- (z-1.5)^2/0.1\right) . 
\end{eqnarray} 
These three examples cover three classes of the possible behavior of $w(z)$. The
first corresponds to a constant equation of state parameter, the second to a
slow transition from a dust like component at large redshifts to a cosmological
constant type at low redshifts and the third corresponds to a sharp transition
in the value of a scalar field occurring around redshift $z = 1.5$ (see
\cite{2009MmSAI..80..785N} for further discussion).

We emphasize that the first two parametrizations lead to a variation of $\Delta
\alpha/\alpha$ that {\it does not} satisfy current geophysical bounds of the
Oklo natural reactor \cite{Gould:2007au,Petrov:2005pu}. (Although the
interpretation of the Oklo bound is not as straightforward as one may think, this
is not the place to discuss it.) Thus for the purposes of the present paper we
use these 3 parametrizations solely as a proof-of-concept representatives of the
whole zoo of models, for the purpose of exploring the reconstruction method.

\section{\label{results}Exploring the pipeline} 

We are now in a position to start a forecast  analysis of the dark energy equation of state for our three fiducial forms of $w^F(z)$. 
We take a total number of bins between redshift 0 and 4 to be 30. We assume a
sample of 3000 supernovae distributed between redshift 0 and 1.7 (with 13 bins)
with an uncertainty on the magnitude of $\sigma_m = 0.11$. These numbers are typical 
of future supernovae datasets.
For the spectroscopic
measurements we use a distribution between redshift 0.5 and 4 (with 27 bins) and
we will consider three different scenarios. Some bins overlap so we obtain in total
30 bins.

We will assume a flat universe, and further simplify the analysis by fixing $\Omega_m=0.3$. This is a
standard procedure, that was followed in the original paper of Huterer and
Starkman  \cite{Huterer:2002hy} and also in a number of subsequent works. The goal of the
analysis is to characterize gains in sensitivity as future, more precise
datasets become available, rather than provide hard numbers. Therefore,
although this is certainly a simplifying assumption, it is a legitimate
one. This specific choice of $\Omega_m$ has a negligible effect on the main
result of the analysis, which is the uncertainty in the best determined
modes.

\subsection{Forthcoming datasets}

Based on current plans for ESPRESSO \footnote{See http://espresso.astro.up.pt/}
and CODEX \footnote{http://www.iac.es/proyecto/codex/} (see also Refs.~\cite{Cristiani:2007by,Liske:2008zu}), we will consider three different simulated datasets of 
fine-structure constant measurements:

\begin{itemize}
\item A \textbf{baseline scenario}, in which we will assume 30 systems with
$\sigma_{\Delta\alpha/\alpha} = 6\times10^{-7}$ for ESPRESSO and 100 systems
with $\sigma_{\Delta\alpha/\alpha} = 1\times10^{-7}$ for CODEX, uniformly
distributed in the redshift range $0.5<z<4$. This is meant to represent what we
can confidently expect to achieve in a relatively short amount of time once the spectrographs are operational
(within 3 to 5 years of data acquisition), given the
current plans for their sensitivity, and it will therefore provide the basis for most of our discussion.
\item An \textbf{ideal scenario}, in which we will assume 100 systems with
$\sigma_{\Delta\alpha/\alpha} = 2\times10^{-7}$ for ESPRESSO and 150 systems
with $\sigma_{\Delta\alpha/\alpha} = 3\times10^{-8}$ for CODEX. This is optimistic both in the uncertainty of individual measurements and in the number
of measurements. Although several hundred absorbers are already known where
these measurements can be carried out, the sources are quite faint and putting
together such a dataset would at the very least require a very long time. Having
said that, our goal in considering this case is to obtain an indication for the
dependence of our results on the uncertainty and number of the measurements.
\item A \textbf{control scenario}, which is meant to be somewhat more realistic
from an observational perspective in the sense that we do not assume the same
uncertainty for all measurements (although we still assume that they are
uniformly distributed in the redshift range $0.5<z<4$). In this case, for
ESPRESSO, we assumed the same number of sources as in the baseline scenario but
with the uncertainties drawn from a normal distribution centered on
$\sigma_{\Delta\alpha/\alpha} = 6 \times 10^{-7}$ and with standard deviation 
$\sigma_{\Delta\alpha/\alpha}/2$; the same can also be done for CODEX. This is a
computationally simple way to check how the pipeline handles non-uniform
uncertainties and is also a proxy for the effect of redshift coverage -- an
important issue when defining an observational strategy.
\end{itemize}

\begin{figure}
\includegraphics[width=8cm]{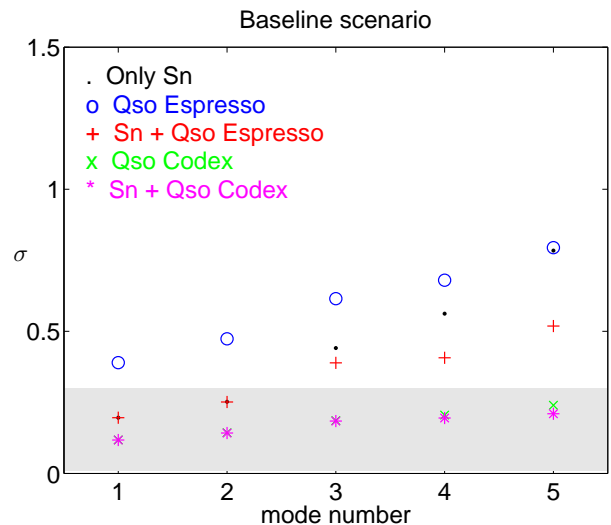}
\caption{\label{sigma2baselineP1} The error $\sigma_i$ for the five best determined modes for the fiducial parametrization (\ref{fid1}) in the baseline scenario.}
\end{figure}

\begin{figure}
\includegraphics[width=8cm]{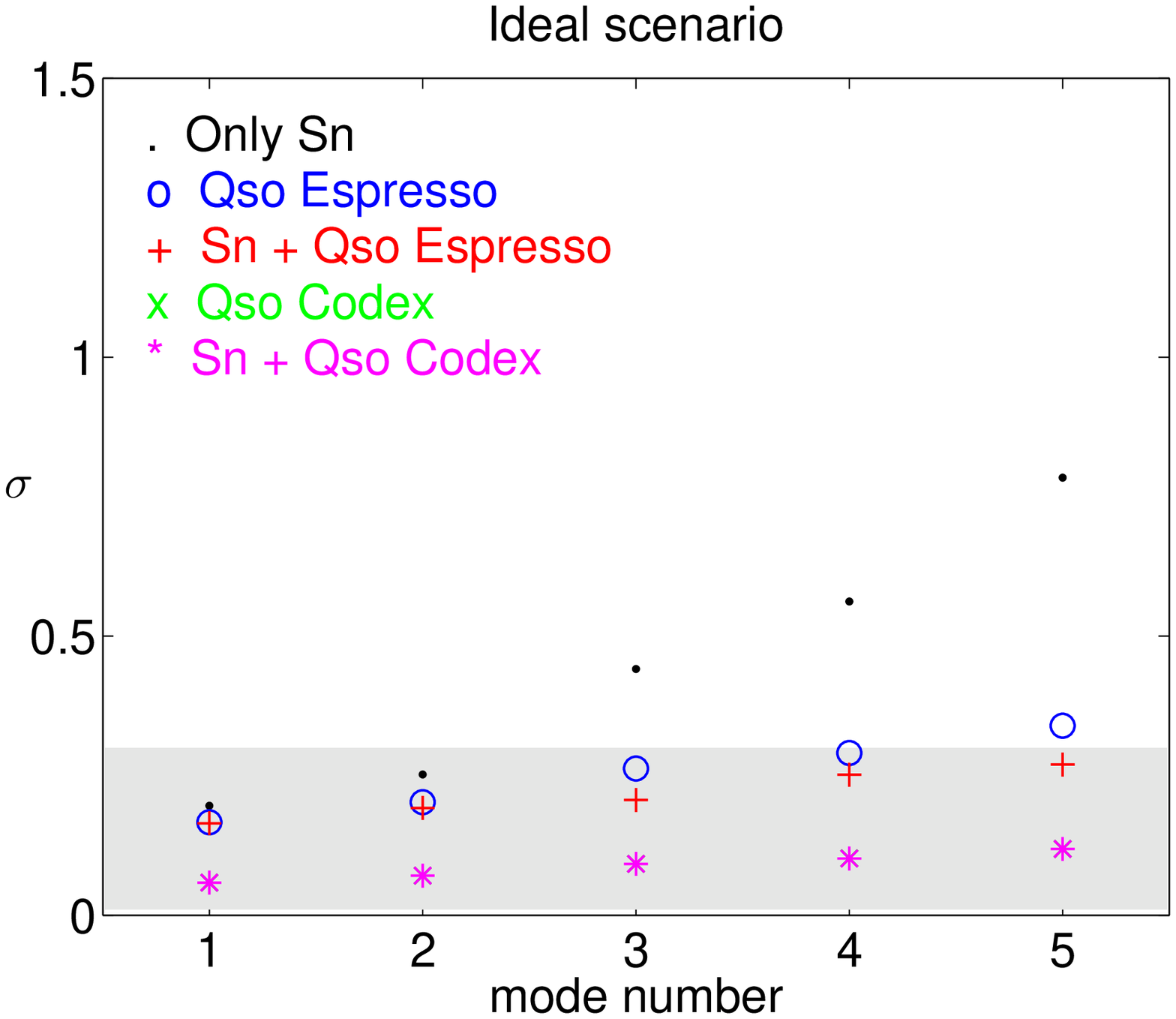}
\caption{\label{sigma2idealP1} The error $\sigma_i$ for the five best determined modes for the fiducial parametrization (\ref{fid1}) in the ideal scenario.}
\end{figure}

\begin{figure}
\includegraphics[width=8cm]{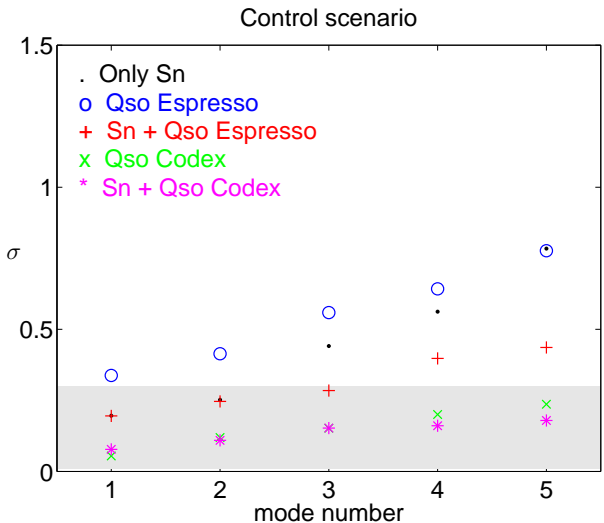}
\caption{\label{sigma2controlP1} The error $\sigma_i$ for the five best determined modes for the fiducial parametrization (\ref{fid1}) in the control scenario.}
\end{figure}

We can now use the principal component approach to compare the various data sets as probes of dark energy. We start by showing the spectra of eigenvalues (or equivalently, the error on the mode $i$, $\sigma_i^2 = 1/\lambda_i$) for the different scenarios that we proposed to analyse, for parametrization  (\ref{fid1}). This is illustrated in Figs.~\ref{sigma2baselineP1}, \ref{sigma2idealP1} and \ref{sigma2controlP1}. 
The shaded region represents a rough threshold $\sigma = 0.3$ below which modes are considered well determined, and therefore, informative \cite{Crittenden:2005wj}.
By inspecting Figs.~\ref{sigma2baselineP1} to 
\ref{sigma2controlP1} it can easily be understood that the baseline and control scenarios yield identical results for the magnitude of the modes' errors. The ideal scenario, as expected, permits that more eigenmodes enter the region for which it is considered  that modes are informative. In order to avoid an overburden of similar figures, we summarize the eigenvalue spectra for this parametrization (\ref{fid1}) and parametrizations (\ref{fid2}) and (\ref{fid3}) 
in Tables \ref{tabela1} -- \ref{tabela3}. It can be noticed that the baseline and control scenarios for ESPRESSO cannot bring further information when compared with supernovae alone. Only in the ideal scenario {\it and} in combination with supernovae, can the number of useful modes raise above 2. The only exception is for parametrization (\ref{fid1}) where the ideal ESPRESSO has 4 good modes. CODEX, however, will offer a set of high precision data which can on its own provide a large number of informative eigenmodes in either of the three scenarios. 
We can see that the results are fairly independent of the fiducial parametrization chosen though parametrization (\ref{fid1}) allows, in general, a higher number of modes considered informative for a given combination of data.
\begin{table}
\begin{tabular}{|c|c|c|c|}
\hline
 ~ & Beseline & Ideal & Control \\ 
 \hline
Sne & 2 & 2 & 2 \\
\hline
ESPRESSO & 0 & 4 & 0 \\
\hline
Sne + ESPRESSO & 2 & 5 & 3 \\
\hline
CODEX & 7 & 16 & 7 \\
\hline
Sne + CODEX & 9 & 18 & 10 \\
\hline
\end{tabular}
\caption{\label{tabela1} The table indicates, for the fiducial parametrization (\ref{fid1}), how many modes have an error below the threshold value $\sigma = 0.3$ considering the baseline, ideal and control scenarios.}
\end{table}
\begin{table}
\begin{tabular}{|c|c|c|c|}
\hline
 ~ & Beseline & Ideal & Control \\ 
 \hline
Sne & 2 & 2 & 2 \\
\hline
ESPRESSO & 0 & 1 & 0 \\
\hline
Sne + ESPRESSO & 2 & 3 & 2 \\
\hline
CODEX & 2 & 8 & 3 \\
\hline
Sne + CODEX & 4 & 10 & 5 \\
\hline
\end{tabular}
\caption{\label{tabela2} The table indicates, for the fiducial parametrization (\ref{fid2}), how many modes have an error below the threshold value $\sigma = 0.3$ considering the baseline, ideal and control scenarios.}
\end{table}
\begin{table}
\begin{tabular}{|c|c|c|c|}
\hline
 ~ & Beseline & Ideal & Control \\ 
 \hline
Sne & 2 & 2 & 2 \\
\hline
ESPRESSO & 0 & 2 & 0 \\
\hline
Sne + ESPRESSO & 2 & 4 & 2 \\
\hline
CODEX & 4 & 12 & 6 \\
\hline
Sne + CODEX & 6 & 14 & 7 \\
\hline
\end{tabular}
\caption{\label{tabela3} The table indicates, for the fiducial parametrization (\ref{fid3}), how many modes have an error below the threshold value $\sigma = 0.3$ considering the baseline, ideal and control scenarios.}
\end{table}

Let us now look at the evolution of the eigenmodes for parametrization (\ref{fid1}) and compare with the modes using supernovae data alone.
\begin{figure}
\includegraphics[width=8cm]{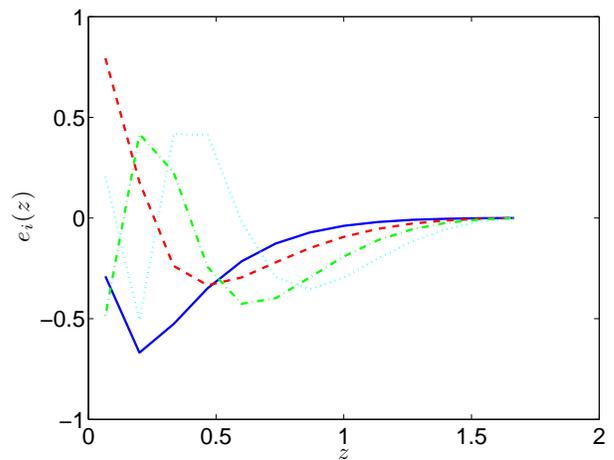}
\caption{\label{eiSneP1} The four best determined eigenmodes for parametrization (\ref{fid1}) using only supernovae. Solid line, dashed line, dash-dotted line and dotted line correspond to first, second, third and fourth modes, respectively.}
\end{figure}
Because only the first two modes are considered informative, a reconstruction of the equation of state parameter using only supernovae can only be made reliably up to redshift 0.5 using the principal component approach. This value corresponds roughly to the position of the second mode's peak in Fig.~\ref{eiSneP1}. We are interested in evaluating whether this limit can be relaxed with the help of QSO data using the ESPRESSO and CODEX spectrographs.
\begin{figure}
\includegraphics[width=8cm]{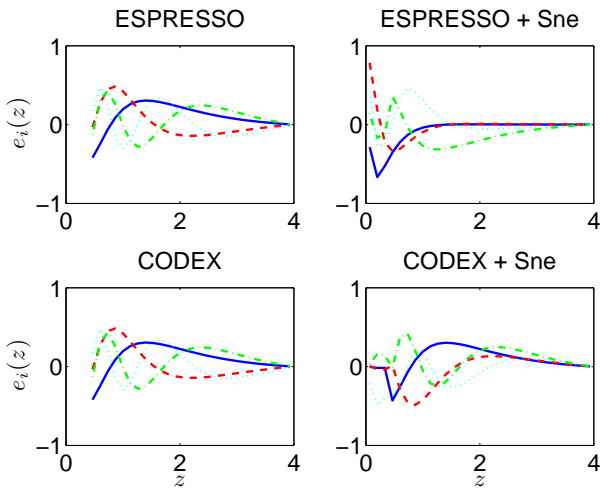}
\caption{\label{eitejo1P1} The four best determined eigenmodes using parametrization (\ref{fid1}) for the baseline scenario. Solid line, dashed line, dash-dotted line and dotted line correspond to first, second, third and fourth modes, respectively.}
\end{figure}
For the baseline scenario we see that only CODEX will permit a large number of modes considered useful. Inspecting Fig.~\ref{eitejo1P1} we see that a reconstruction of the equation of state parameter could be in principle possible up to redshift 2 corresponding roughly to the third peak of the third mode. The conclusions are similar for the ideal and control scenarios, see Figs.~\ref{eitejo2P1} and \ref{eitejo3P1}. In particular, only for the ideal scenario would ESPRESSO add information on the nature of dark energy.
\begin{figure}
\includegraphics[width=8cm]{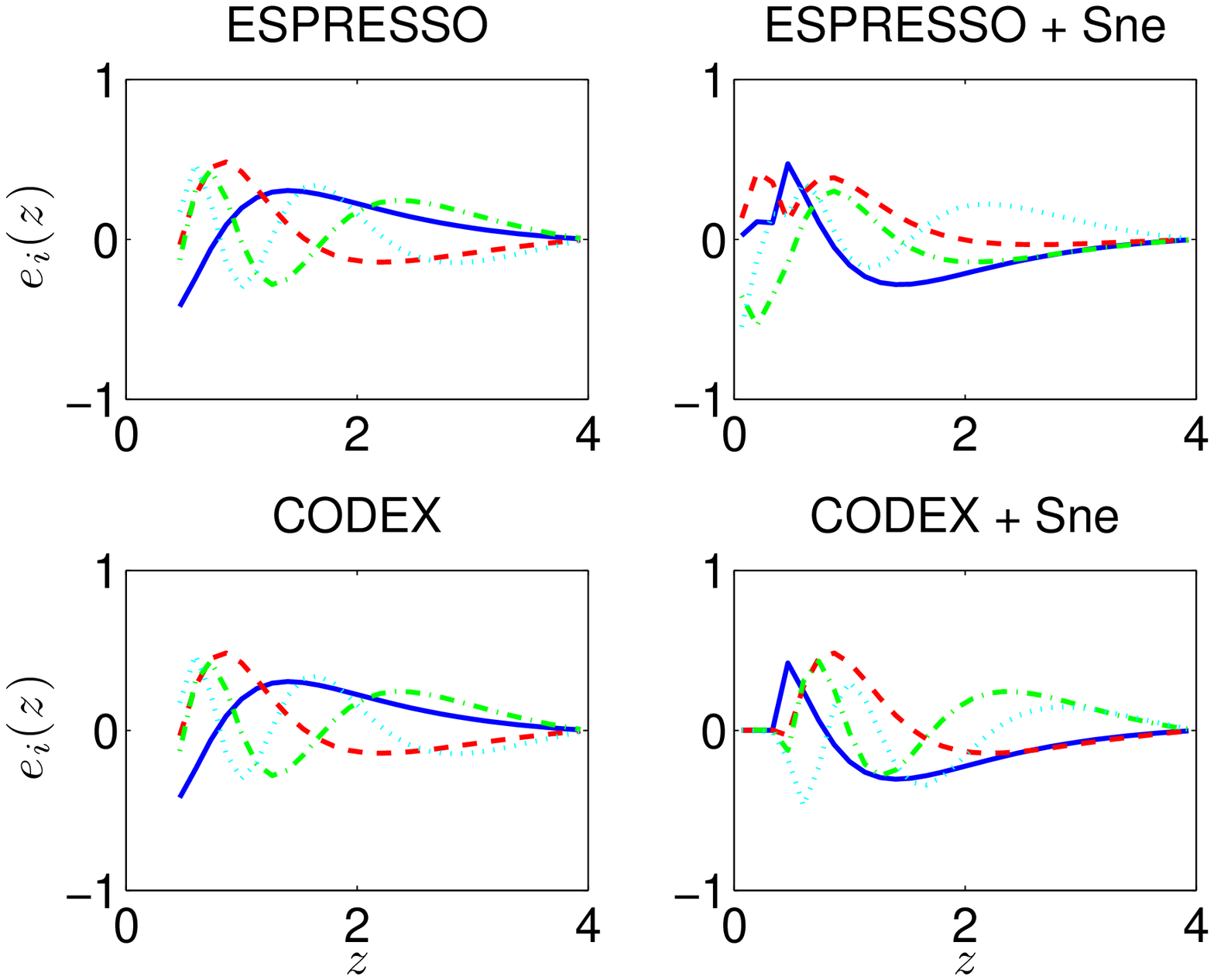}
\caption{\label{eitejo2P1} The four best determined eigenmodes using parametrization (\ref{fid1}) for the 
ideal scenario. Solid line, dashed line, dash-dotted line and dotted line correspond to first, second, third and fourth modes, respectively.}
\end{figure}
\begin{figure}
\includegraphics[width=8cm]{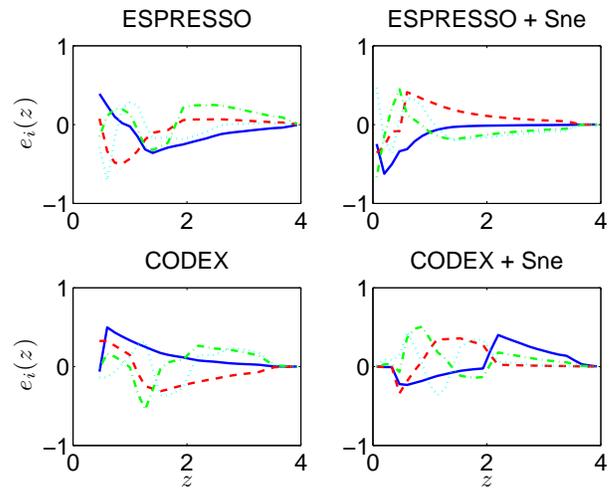}
\caption{\label{eitejo3P1} The four best determined eigenmodes using parametrization (\ref{fid1}) for the 
control scenario. Solid line, dashed line, dash-dotted line and dotted line correspond to first, second, third and fourth modes, respectively.}
\end{figure}
The form of the fiducial function $w^F(z)$ is not determinant in the results and we would have reached the same conclusions had we illustrated the eigenmode evolution of parametrizations (\ref{fid2}) or (\ref{fid3}). We must emphasize, however, that the choice of threshold does play a role. Had we chosen a slightly higher value, say $\sigma = 0.5$ then the baseline scenario of ESPRESSO could indeed bring new information on the nature of dark energy when comparing to supernovae observations.

\section{Reconstructing the equation of state parameter $w(z)$}
\label{reconstruction}
Folowing Refs.~\cite{Huterer:2002hy,2007PhRvD..75j3003A} one can now attempt a reconstruction $w(z)$ by keeping only the most accurately determined modes (the ones with largest eigenvalues). To do this, we need to decide how many components to keep. We must point out that the weak point of this procedure consists in neglecting the high frequency modes. In a more recent analysis \cite{Crittenden:2005wj} and perhaps more robust approach, all modes are kept and a correlation function describing fluctuations from a fiducial model is chosen. This method allows a more accurate reconstruction of the dark energy parameter over the whole range of redshifts covered by observations. We decided, however, to follow Refs.~\cite{Huterer:2002hy,2007PhRvD..75j3003A} given the simplicity of the methods described there. This will be sufficient to support our main conclusion i.e., that a combination of supernovae and quasar data will improve on testing the time dependence of dark energy.

\subsection{Selection of components: risk vs. normalization}

One may argue that the optimal value of modes $M$ to be kept corresponds to the value that minimizes the risk, defined as \cite{Huterer:2002hy}
\begin{equation}
risk = bias^2 + variance ,
\end{equation}
with 
\begin{equation} 
bias^2(M) = \sum_{i=1}^N\left( \tilde w(z_i) - w^F(z_i) \right)^2 ,
\end{equation} 
where the notation $\tilde w$ means that the sum in (\ref{recw}) runs from 1 to $M$, and 
\begin{equation} 
variance = \sum_{i=1}^N \sum_{j=1}^M \sigma^2(\alpha_j) 
e_j(z_i). 
\end{equation} 
The bias measures how much the reconstructed equation of state, $w_{\rm
rec}(z)$, differs from the true one by neglecting the high and noisy modes and
therefore, typically decreases as we increase $M$. The variance of $w(z)$,
however, will increase as we increase $M$, since we will be including modes that
are less accurately determined.

An alternative way to decide on the number of optimal modes is to choose the
largest value for which the error is below unity, or equivalently, the RMS
fluctuations of the equation of state parameter in such mode are
\begin{equation}
\langle (1+w(z))^2 \rangle = \sigma_i^2 \lesssim 1\,.
\end{equation}
Having thus determined the optimal number of modes, we proceed with the
normalization of the error following Ref.~\cite{Albrecht:2009ct} such that
$\sigma^2 = 1$ for the worse determined mode and normalize the error on the
remaining modes by taking
\begin{equation}
\sigma^2(\alpha_i) \rightarrow \sigma_n^2(\alpha_i) =
\frac{\sigma^2(\alpha_i)}{1+\sigma^2(\alpha_i)} .
\end{equation}

The PCA allows us to optimize an experiment towards the range in redshift we are
interested in. In our work, we will use a combination of supernovae and quasar
absorption lines to understand how well the equation of state parameter $w(z)$
will be  constrained with forthcoming data on cosmological variation of
fundamental parameters obtained with the spectrographs ESPRESSO for the VLT and
CODEX for the E-ELT.

\begin{figure}
\includegraphics[width=8cm]{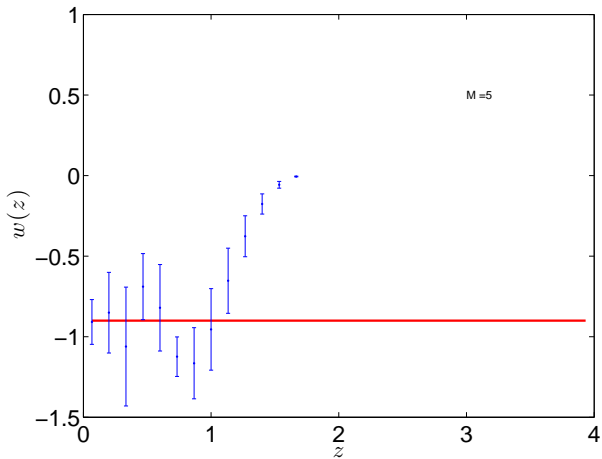}
\caption{\label{wSneP1} Reconstruction of the equation of state parameter (\ref{fid1})  using only supernovae with the minimization of the risk method.}
\end{figure}
%

\begin{figure}
\includegraphics[width=8cm]{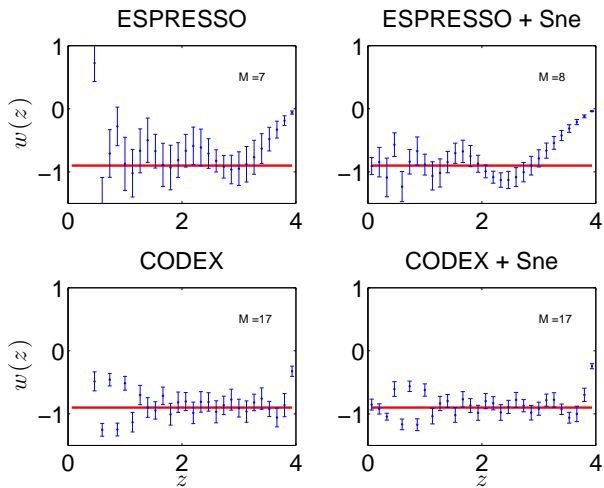}
\caption{\label{wtejo1P1} Reconstruction of the equation of state parameter (\ref{fid1}) in the baseline scenario with the minimization of the risk method.}
\end{figure}
\begin{figure}
\includegraphics[width=8cm]{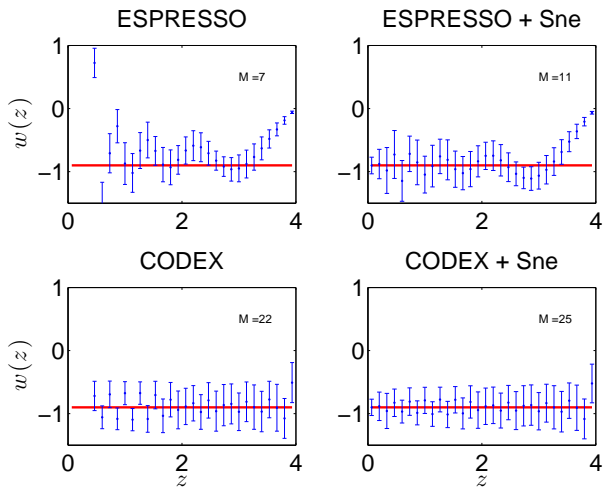}
\caption{\label{wtejo1P1N} Reconstruction of the equation of state parameter (\ref{fid1}) in the baseline scenario with the normalization of the error on the modes method.}
\end{figure}

In Fig.~\ref{wSneP1} we illustrate the reconstruction using only supernovae for our fiducial model (\ref{fid1}) and in Figs.~\ref{wtejo1P1} and \ref{wtejo1P1N} we show the result of the reconstruction for the two spectrographs and the two
methods for the selection of the number of components.
We can observe that, for this fixed number of bins, the reconstruction obtained
using supernovae is only accurate up to redshift $\approx 1$. In particular, because we neglect the poorly determined modes
which are the ones with high amplitudes for bins of large redshift, the
reconstructed equation of state parameter tends to zero for large redshift. This unavoidable feature of the PCA truncation method can be confused with a real increase in the equation of state at high redshift. As it was pointed out in \cite{Huterer:2002hy}, had we reconstructed $1+w(z)$ instead, then $w(z)$ would approach $-1$ for large redshifts. Ideally we would like to extend the survey to large redshifts but unfortunately we only expect data from supernovas up to redshift $z \approx 1.7$.

Measurements from the quasar absorption lines, which are available for a larger
redshift interval, provide in general a more reliable reconstruction. For our
fiducial parametrizations of $w(z)$, these datasets can give a qualitatively
accurate account of the evolution of the equation of state parameter to fairly
high redshifts.

Comparing the various fiducial models for the same observational dataset shows
that (as one would expect) the redshift up to which the reconstruction remains
accurate depends in part on the correct underlying model, specifically on whether
its equation of state remains close to a cosmological constant or approaches a
dust-like behavior. However, comparing the CODEX and ESPRESSO cases show that
one can go deeper in redshift by increasing the sensitivity of the measurements,
since that allows one to add components to the reconstruction. 
 In particular, the truncation problem mentioned earlier becomes less problematic as the reconstructed equation of state parameter no longer approaches zero for large redshifts. This is a true statement also for the alternative parametrizations.

The combination of supernovae with quasar absorption lines data further improves
the determination of the equation of state parameter. In particular, we can now
obtain information on $w(z)$ all the way from $z \approx 0$ up to $z \approx 3$.
 The reconstruction using CODEX, benefiting from an almost one order of
magnitude improvement in the sensitivity of the QSO data points, is
substantially better than the one obtained with ESPRESSO. 

We can also compare the two methods of determining the optimal value of modes to
keep in the reconstruction. We have seen before that the minimization of the
risk method is a compromise between having an accurate equation of state
reconstruction and having a small error bar in this reconstruction. The
normalization of the error on the modes method, however, makes use of our prior
prejudice that variations of $w(z)$ larger than unity are unlikely. We observe,
from comparing Fig.~\ref{wtejo1P1} to Fig.~ \ref{wtejo1P1N}, to which correspond different methods, that the latter method picks
more modes, which leads to a more accurate reconstruction. Since we are
including additional modes with progressively larger errors, the reconstructed
equation of state in this case also has larger error bars. In other words, the
normalization method provides a more conservative and accurate approach, while
the risk method provides (appropriately) a more aggressive approach.

From a practical point of view, the advantage of PCA techniques is that they may provide us with a simple computational tool to continually optimize an observational plan. In other words, given an ongoing observational campaign in which one has already observed a certain number of sources (with given uncertainties) one can use these tools to simulate improved datasets with the goal of determining how to best spend the remaining telescope time (reducing the uncertainty in measurements of particular sources) so as to achieve the best
possible constraints on these models, given any relevant observational
limitations.

\section{\label{concls} Conclusions}

In the first part of our study we considered the threshold value presented in Ref.~\cite{Crittenden:2005wj} as a mean to determine how many modes can be considered good and informative in describing dark energy. This value, $\sigma \leq 0.3$, is somehow arbitrary but serves our purpose to indicate that future precise measurements of the fine-structure constant, specially with CODEX, can add information on the nature of dark energy up to high redshifts when comparing to supernovae measurements.

In the second part of this work we have compared reconstructions of the equation of state parameter
$w(z)$ of dark energy using a principal component analysis. To this effect we
used a combination of expected supernovae data and quasar absorption
measurements of the fine-structure constant $\alpha$, expected to be available
with the spectrographs ESPRESSO and CODEX. We considered several possible
datasets and also two methods of choosing the best determined modes of the
principal decomposition and studied the effect of the size of the error bars on
the reconstruction.

Our analysis indicates that the normalization of the error on the modes method
appears to give more accurate (closer to the fiducial value) but less precise (more
conservative errors) reconstructions with respect to the risk minimization procedure.
 We also conclude that a reconstruction using
quasar absorption lines is expected to be more accurate than using supernovae
data. However, since the two types of measurements probe different (but
overlapping) redshift ranges, combining them leads to a more complete picture of
the evolution of the equation of state parameter between redshift zero and four.

 A natural extension of this work is to include in addition cosmological measurements of $\mu = m_p/m_e$, the ratio of the proton and electron masses \cite{Thompson}.  Measurements of $\mu$ will be
fewer than those of $\alpha$ but will all be at high redshift and fairly
precise and are, therefore, expected to reduce the errors on the reconstruction of  $w(z)$.

Although in this work we have focused on combining two datasets, we should also
point out that they can also be used separately to provide independent
reconstructions. Comparing the two reconstructions will then provide a
consistency test, specifically for the assumption on the coupling between the
scalar field and electromagnetism (given by Eq.~\ref{coupling}). If so one can
also obtain a measurement for the coupling parameter $\zeta$. A more detailed
treatment of this case, as well as the application of the method to existing
datasets, is left for forthcoming work.

Since a PCA reconstruction involves a truncation, it will not be precise at high redshifts.  How much of a problem this is depends on the behaviour of the true equation of state: the reconstructed equation will always approach $0$ at high redshifts, while the true one may or may not do so in the redshift range being probed by the detaset under consideration. A relevant question for PCA studies is therefore how deep in redshift can one confidently go. In most previous works one has a fiducial model for $w(z)$ that approaches $0$ by around $z\sim1$, in which case supernovas perform well; however this is an optimistic assumption that need not be true. One of our key points is that, not knowing a priori what the correct equation of state will look like, there is strong interest in trying to reconstruct as deep as possible in redshift, and varying couplings are a possible way of doing that.

One must bear in mind that one should not compare the parametric-free PCA constraints with those obtained using parametrized equations of state. Perhaps a more direct comparison is with redshift binned $w(z)$. A good benchmark can be found in Fig. 17 of the work by Kowalsky \textit{et al.} \cite{Kowalski:2008ez}, where one sees how poorly the high redshift behaviour of the equation of state is measured by current data.

Finally, our main conclusion is that, when one compares like with like, the inclusion of varying constants data allows a reliable
reconstruction to be carried out to significantly higher redshifts. Admittedly, for these methods to be competitive will require very good quality measurements, which do not presently exist (although they are expected to be available in a few years), but this point applies both for the supernovas and for the varying couplings. In any case, the method is interesting on its own right, and should be further studied as the science cases of future facilities is developed.

\begin{acknowledgments} 
This work was done in the context of the FCT-DAAD cooperation grant 'The Dark
Side of the Universe' (reference 441.00 Alemanha), with additional support from
project PTDC/FIS/111725/2009 from FCT, Portugal.

The work of CM is funded by a Ci\^encia2007 Research Contract, funded by
FCT/MCTES (Portugal) and POPH/FSE (EC), and is also partially supported by grant
PTDC/CTE-AST/098604/2008. LA and NJN are supported by Deutsche Forschungsgemeinschaft (project TRR33), and NJN is also partially supported by grants CERN/FP/116398/2010 and PTDC/FIS/102742/2008. The work of PP was partially funded by grant
CAUP-09/2009-BII. 

CM acknowledges useful discussions with Paolo Molaro and Patrick Petitjean,
particularly for defining the observational scenarios considered.
\end{acknowledgments}

\bibliography{pca}

\end{document}